\documentclass[fleqn,twoside]{article}
\usepackage{espcrc2}
\usepackage{amssymb}

\usepackage{epsfig}
\usepackage[figuresright]{rotating}

\newcommand{\AmS}{{\protect\the\textfont2
  A\kern-.1667em\lower.5ex\hbox{M}\kern-.125emS}}
\newcommand{\beq}{\begin{equation}}
\newcommand{\eeq}{\end{equation}}
\newcommand{\beqr}{\begin{eqnarray} \nonumber}
\newcommand{\eeqr}{\end{eqnarray}}
\newcommand{\beqrb}{\begin{eqnarray}}
\newcommand{\eeqrb}{\nonumber \end{eqnarray}}

\newcommand{\unit}[1]{\hat{\mathbf{#1}}}

\newcommand{\cm}{\mbox{ cm}}
\newcommand{\sr}{\mbox{ sr}}
\newcommand{\se}{\mbox{ s}}
\newcommand{\yr}{\mbox{ yr}}
\newcommand{\erg}{\mbox{ erg}}
\newcommand{\Hz}{\mbox{ Hz}}
\newcommand{\kHz}{\mbox{ kHz}}
\newcommand{\MHz}{\mbox{ MHz}}
\newcommand{\GHz}{\mbox{ GHz}}

\newcommand{\km}{\mbox{ km}}

\newcommand{\Mpc}{\mbox{ Mpc}}
\newcommand{\eV}{\mbox{ eV}}
\newcommand{\keV}{\mbox{ keV}}

\newcommand{\TeV}{\mbox{ TeV}}
\newcommand{\mK}{\mbox{ mK}}

\newcommand{\K}{\mbox{ K}}
\newcommand{\de}{^{\circ}}

\newcommand{\muG}{\mbox{ $\mu$G}}
\newcommand{\Jy}{\mbox{ Jy}}
\newcommand{\mJy}{\mbox{ mJy}}
\newcommand{\muJy}{\,\mu\mbox{Jy}}
\newcommand{\fin}{\mbox{ .}}
\newcommand{\coma}{\mbox{ ,}}

\newcommand{\gama}{$\gamma$}
\newcommand{\HI}{$\mbox{H \small I}$ }
\newcommand{\HII}{$\mbox{H \small II}$ }

\newcommand{\IUnits}{\erg\se^{-1}\cm^{-2}\sr^{-1}\Hz^{-1} }

\newcommand{\ICUnits}{\keV\se^{-1}\cm^{-2}\sr^{-1} }

\newcommand{\LCDM}{$\Lambda$CDM }
\newcommand{\Lya}{Ly$\alpha$ }

\newcommand{\mla}{\lesssim}
\newcommand{\mga}{\gtrsim}
\newcommand{\marcmin}{^\prime}

\newcommand{\nospaced}{\!\!\!\!} 
 
\newcommand{\nospacef}{\!\!\!\!\!\!} 
\newcommand{\nospaceg}{\!\!\!\!\!\!\!} 

\title{Searching for Intergalactic Shocks with the SKA}

\author{
U. Keshet\address{Physics Faculty, Weizmann Institute, Rehovot 76100, Israel},
E. Waxman\addressmark, and A. Loeb\address{Astronomy Department, Harvard University, 60 Garden Street, Cambridge, MA 02138, USA}\thanks{Einstein Minerva Center, Physics Faculty, Weizmann Institute of Science}
}
       
\begin{document}

\begin{abstract}
Strong intergalactic shocks are a natural consequence of structure formation in the universe. These shocks are expected to deposit large fractions of their thermal energy in relativistic electrons ($\xi_e\simeq 0.05$ according to supernova remnant observations) and magnetic fields ($\xi_B\simeq 0.01$ according to cluster halo observations). We discuss the expected synchrotron emission from such shocks, and the observational consequences for next generation radio telescopes such as the Square Kilometer Array. We present an analytical model, calibrated and verified based on a hydrodynamical \LCDM simulation. The resulting signal composes a large fraction (up to a few $10\%$) of the extragalactic radio background below $500 \MHz$. The associated angular fluctuations, e.g. $\delta T_l \mga 260 (\xi_e\xi_B/5\times 10^{-4}) (\nu/100\MHz)^{-3}\K$ for multipoles $400\mla l\mla 2000$, dominate the radio sky for frequencies $\nu \mla 10 \GHz$ and angular scales $1\marcmin \mla \theta < 1\de$ (after a modest removal of point sources), provided that $\xi_e \,\xi_B\mga 3\times 10^{-4}$. The fluctuating signal is most pronounced for $\nu \mla 500 \MHz$, dominating the sky there even for $\xi_e\,\xi_B = 5\times 10^{-5}$. We find the signal easily observable by next generation telescopes such as the SKA, and marginally observable with present telescopes. The signal may be identified using cross-correlations with tracers of large-scale structure (e.g. \gama-ray emission from intergalactic shocks). Detection of the signal will provide the first identification of intergalactic shocks and of the warm-hot intergalactic medium (believed to contain most of the baryons in the low redshift universe), and gauge the unknown intergalactic magnetic field.
\vspace{1pc}
\end{abstract}

\maketitle

\section{Introduction}

\label{sec:Introduction}
\indent
The gravitational formation of structure in the universe inevitably produced strong, collisionless shocks in the intergalactic medium (IGM), owing to the convergence of large-scale flows. In these shocks, electrons are believed to be Fermi accelerated up to highly relativistic ($\mga 10 \TeV$) energies, limited by inverse-Compton cooling off cosmic microwave background (CMB) photons \cite{LoebWaxman2000}. The resulting \gama-ray emission traces the large-scale structure of the universe. Rich galaxy clusters should be detected by future \gama-ray missions as \gama-ray sources \cite{LoebWaxman2000,Totani00,WaxmanLoeb2000} in the form of accretion rings with bright spots at the intersections with galaxy filaments \cite{Keshet03}. Recently, a possible association of $\gamma$-rays (measured by the \emph{Energetic Gamma-Ray Experiment Telescope}, EGRET) with Abell cluster locations was identified at a $3\sigma$ confidence level 
(\cite{Scharf2002}, but see also \cite{Reimer03}). 

In addition to the inverse-Compton emission from intergalactic shocks,
synchrotron radiation should also be emitted by the relativistic electrons, as they gyrate in the shock-induced magnetic fields. The resulting radio signature is expected to trace the structure of the universe at low redshifts ($z\mla1$), and to be dominated by rich, young galaxy clusters. Indeed, extended radio emission with no optical counterpart is observed in about $10\%$ of the rich galaxy clusters (radio halos and radio relics, see \cite{Giovannini99}), and in more than a third of the young, massive clusters (with X-ray luminosity $L_X>10^{45} \erg \se^{-1}$, see \cite{Feretti03}). The radio emission has been identified as synchrotron radiation from relativistic electrons, but there are different models for the origin of the electrons and the magnetic fields involved (for a recent discussion, see \cite{Bagchi03}). 
Observations of nine radio relics were used \cite{Ensslin98} to suggest an association between these sources and structure formation shocks, focusing on the possibility that the relics are revived fossil radio cocoons originating from nearby radio galaxies, re-energized through diffusive shock acceleration. Recently, large-scale diffuse radio emission was discovered around a filament of galaxies, possibly tracing an accretion shock on this scale \cite{Bagchi_etal02}.

Waxman \& Loeb (2000, \cite{WaxmanLoeb2000}) have proposed a simple model, which allows one to estimate both the radio and the \gama-ray signatures of intergalactic shocks, produced by electrons accelerated from the inflowing plasma. Their model allows one to estimate the radio and the \gama-ray backgrounds and their anisotropy characteristics, as well as the signature of individual clusters. The model uses dimensional analysis arguments to estimate the properties of the virialization accretion shock of a halo, as a function of redshift $z$ and halo mass $M$. Halo abundance estimates at different redshifts [such as the Press \& Schechter (1974, \cite{PressSchechter}) halo mass function], may then be used to calculate various observables. The model approximates the strong accretion shocks as being spherically symmetric, and neglects weak merger shocks. A fraction $\xi_e \simeq 0.05$ of the shock thermal energy is assumed to be carried by relativistic electrons, based on observations of supernovae remnants (SNRs, for a discussion, see \cite{Keshet03}). Observations of $\mga 0.1 \muG$ magnetic fields in cluster halos (for a recent review, see \cite{Carilli02}) require that a fraction $\xi_B\simeq 0.01$ of the shock thermal energy be transferred into downstream magnetic fields. The model has been generalized for a \LCDM universe by Keshet, Waxman \& Loeb (2004a, \cite{Keshet04a}). The free parameters of the model were calibrated according to global features of a \LCDM simulation, and verified to give radio and \gama-ray signatures consistent with the signatures found directly from the simulation. Shock asymmetry was incorporated into the model by introducing a geometrical correction factor, approximately measured from the simulation. 

Identification of radio or \gama-ray emission from intergalactic shocks
holds a great promise for advancing current knowledge on shock formation in
the IGM. It should provide the first direct evidence for such shocks,
revealing the underlying large-scale cosmological flows. When combined with
\gama-ray detection, the radio signal will provide a direct measure of the
unknown magnetic fields in the IGM, possibly shedding light on the processes leading to IGM magnetization. Emission from large-scale shocks traces the
undetected warm-hot IGM at temperatures $10^5\K \mla T\mla 10^7 \K$ that is
believed to contain most of the baryons in the low redshift universe
(see e.g. \cite{Davea01}). Moreover, the signal may be used to study
non-thermal physical processes in the intergalactic environment, such as
Fermi acceleration in low density shocks.

In this chapter we discuss the radio signature of intergalactic shocks, and the observational consequences for next generation radio telescopes such as the Square Kilometer Array. 
In \S\ref{sec:IGM_model} we review the model for radio emission from intergalactic shocks, and test its validity. In \S\ref{sec:feasibility} we discuss the observational consequences of the predicted radio signal, with emphasis for the SKA, and examine confusion with various foreground and background signals. Finally, \S\ref{sec:discussion} summarizes our results and addresses their potential implications. We discuss methods by which the signal may be identified, and present some consequences of a future positive detection of the signal.


\section{Model} 
\label{sec:IGM_model} 
\indent 

Here we study the extragalactic radio signal expected from the strong
intergalactic shocks involved in structure formation. We begin by constructing the model from dimensional-analysis arguments \cite{WaxmanLoeb2000} in \S \ref{subsec:model_review}. Next, we present a generalized version of the model \cite{Keshet04a}, corrected and calibrated for a \LCDM universe. The energy fractions $\xi_e$ and $\xi_B$ are evaluated in \S \ref{subsec:model_efficiency}, and a comparison between model predictions and radio halo observations is presented in \S \ref{subsec:model_halos}.

In the following we use a 'concordance' $\Lambda\mbox{CDM}$ model: a flat universe with normalized vacuum energy density $\Omega_\Lambda=0.7$, matter energy density $\Omega_M=0.3$, baryon energy density $\Omega_B=0.04$, Hubble parameter $h=0.67$, and an initial perturbation spectrum of slope $n=1$ and normalization $\sigma_8=0.9$. The various parameters of the cosmological model are summarized in Table \ref{tab:CosmoParams}. 

\begin{table}[htb]
\caption{Cosmological model parameters. }
\label{tab:CosmoParams}
\begin{tabular}{@{}lll}
\hline
\em Parameter & \em Meaning & \em Value \\
\hline
$h$           & Hubble parameter & 0.67 \\ 
$k$           & Curvature & 0 \\
\hline 
$\Omega_m$    & Matter energy density & 0.30 \\
$\Omega_{dm}$ & Dark matter energy density & 0.26 \\ 
$\Omega_b$    & Baryon energy density & 0.04 \\ 
$\Omega_\Lambda$ & Vacuum energy density & 0.70 \\
$\chi$      & Hydrogen mass fraction & 0.76 \\ 
\hline 
$n$           & Fluctuation spectrum slope & 1 \\
$\sigma_8$    & Spectrum normalization & 0.9 \\ 
\hline
\end{tabular}\\[2pt]
\end{table}

\subsection{Dimensional Analysis}
\label{subsec:model_review}

Dimensional analysis arguments can be used to relate the mass $M$ of a virialized halo, its velocity dispersion $\sigma$, its smooth mass accretion rate $\dot{M}$ through strong shocks, its temperature $T$ and the radius $r_{sh}$ of its strong accretion shock, by \cite{WaxmanLoeb2000} 
\beq M = \frac{\sqrt{2}}{5}\,\frac{\sigma^3(M,z)}{G H(z)} \coma \eeq 
\beqrb \label{eq:model_Mdot} \dot{M}(M,z) \!\!\!\! & = & \!\!\!\! f_{acc}\frac{\sigma^3(M,z)}{G} \\ & \simeq & \!\!\! 2.5\times 10^{-10} f_{acc}\, h_{70}\, a^{-3/2} g(a) \,M \yr^{-1}
\coma \eeqrb 
\beqrb \label{eq:model_T} T(M,z) \!\!\! & = & \!\!\! f_T k_B^{-1} \,\mu \, \sigma^2 (M,z) \\ & \simeq & \!\!\! 1.8 \times 10^7 f_T \,h_{70}^{2/3} a^{-1} g(a)^{2/3} M_{14}^{2/3} \K \coma \eeqrb 
and 
\beqrb \label{eq:model_Rsh} r_{sh}(M,z) & = & f_{r} \frac{\sqrt{2}}{5} \, \frac{\sigma(M,z)}{H(z)} \\ & \simeq & 1.9 f_{r} \, h_{70}^{-2/3} a \, g(a)^{-2/3} M_{14}^{1/3} \Mpc \fin \eeqrb 
Here $H(z) = 100 \,h \, a^{-3/2}\, g(a)\km \se^{-1} \Mpc^{-1}$ is the Hubble
parameter, with $a\equiv(1+z)^{-1}$ and $g(a)=[\Omega_m+\Omega_\Lambda a^3
+ (1-\Omega_m-\Omega_\Lambda)a]^{1/2}$, $h_{70} \equiv h/0.70$, $\mu\simeq 0.65\,m_p$ is the average mass of a particle (including electrons), $M_{14}\equiv M/10^{14}M_\odot$, and $k_B$ is the Boltzmann constant. The free parameters $f_{acc}$, $f_T$ and $f_{r}$ are dimensionless factors
of order unity, which are roughly constant in the redshift range
and cosmological model of interest, and must be calibrated separately. The
definitions of the model parameters are summarized in Table
\ref{tab:ModelParams}, along with their values as calibrated in \S
\ref{subsec:model_LCDM}. 

Note that the above estimates refer to strong shocks only; weak shocks that may result from mergers of comparable mass objects are ignored since they lead to accelerated electron distributions with little energy in highly relativistic electrons (for a discussion see \cite{Keshet03}). Weak merger events may only \emph{enhance} the predicted radio background from intergalactic shocks, and only in very low photon frequencies. 

Collisionless, non-relativistic shocks are known to accelerate electrons to
highly relativistic energies. In \cite{Keshet03} it was shown, that SNR observations suggest that the energy density of relativistic electrons accelerated by a structure formation shock constitutes a fraction $\xi_e \simeq 5\%$ of the thermal energy density behind the shock (up to a factor of $\sim 2$, see also discussion in \S \ref{subsec:model_electron_efficiency}). The maximal energy an electron can be accelerated to is limited by cooling, predominantly through inverse-Compton scattering of background CMB photons, yielding a maximal electron Lorenz factor $\gamma_{max} \simeq 3 \times 10^7$ \cite{LoebWaxman2000}. 

The model focuses on \emph{strong} shocks, which accelerate electrons 
to a power-law distribution of index $p=2$ in the differential number of accelerated electrons per electron energy (equal energy per logarithmic interval of electron energy). The luminosity of a halo of mass $M$ at redshift $z$ due to inverse-Compton scattering of CMB photons may thus be estimated as 
\beqr  \label{eq:halo_L_IC}
\nu L_\nu^{iC} \nospaceg &(&\nospaceg M,z) \\ & = & {1 \over {2 \ln \gamma_{max}}} \left[ {\Omega_b \over \Omega_m} {{\dot{M} (M,z)} \over \mu} \right] \left[ \xi_e {3 \over 2} k_B T(M,z) \right] \nonumber \\ 
& \simeq & 1.1 \times 10^{42} \left( f_{acc} f_T \frac{\xi_e} {0.05} \right) \nonumber \left( h_{70}^{5/3} \frac{7 \Omega_b}{\Omega_m} \right) \nonumber \\ & & \times \, \left[ a^{-5/2} g(a)^{5/3} \right] M_{14}^{5/3} \erg \se^{-1}\fin \eeqr 

Assuming that the energy density of the downstream magnetic field constitutes a fraction $\xi_B\simeq 1\%$ of the downstream thermal energy density, the magnetic field strength is given by 
\beqr B(M,z) \nospaced &=& \nospaced 0.14 \left( \frac{f_T}{f_{r}^2} \frac{\xi_B}{0.01} \right)^{1/2} \left[ h_{70}^{4/3} \left( \frac{7\,\Omega_b}{\Omega_m} \right)^{1/2} \right] \\ & & \times \left[a^{-2} g(a)^{4/3} \right] M_{14}^{1/3} \muG \coma \eeqr 
consistent with observation of galaxy cluster halos (see \S \ref{subsec:model_magnetic_efficiency} for discussion). This yields a synchrotron luminosity 
\beqrb \label{eq:halo_L_syn}
&\nospacef\nu&\nospaceg L_\nu^{syn} (M,z) = {{B(M,z)^2/8\pi} \over u_{cmb}(z)} \nu L_\nu^{iC}(M,z) \nonumber \\ &=&1.9 \times 10^{39} \left( \frac{f_{acc}
f_T^2}{f_{r}^2} \frac{\xi_e}{0.05} \frac{\xi_B}{0.01} \right) h_{70}^{13/3} \\ & & \nonumber \times  \left( \frac{7\Omega_b}{\Omega_m} \right)^2 \left[a^{-5/2} g(a)^{13/3} \right] M_{14}^{7/3} \erg \se^{-1} \coma \eeqrb
where $u_{cmb}$ is the energy density of the CMB. Note that whereas the
inverse-Compton signal depends on the parameter $\xi_B$ only
logarithmically (through the maximal energy attained by the relativistic
electrons), the synchrotron emission scales almost linearly with $\xi_B$. 

Given the number density of halos of a given mass at a given redshift, $dn(z)/dM$, one may integrate equation (\ref{eq:halo_L_IC}) or equation (\ref{eq:halo_L_syn}) to predict the inverse-Compton or the synchrotron background from intergalactic shocks,
\beq \langle \nu L_\nu \rangle = \int dz \, {{c\,dt}\over dz} \int dM {dn(z) \over dM} {{\nu L_\nu(M,z)} \over {4\pi(1+z)^4}} \fin \eeq 
For example, with the Press-Schechter halo mass function and our cosmological model, we find an inverse-Compton background flux $\langle \nu I_\nu^{iC} \rangle = 1.3\, f_{acc}\, f_T\, (\xi_e/0.05) \keV \se^{-1} \cm^{-2} \sr^{-1}$, and a synchrotron background flux $\langle \nu I_\nu^{syn} \rangle = 3.2 \times 10^{-12} f_{acc}\, f_T^2\, f_{r}^{-2}\, (\xi_e/0.05)\, (\xi_B/0.01) \erg \se^{-1} \cm^{-2} sr^{-1}$. 

The model may be used to calculate the two-point correlation function of the radiation emitted by intergalactic shocks. The low optical depth of large clusters, which dominate the background, enables one to neglect cases where more than one cluster lies along the the line of sight, and approximate \cite{WaxmanLoeb2000} 
\beqrb \label{eq:self_corr} \delta^2 I_\nu (\psi) & \equiv & \langle I_\nu(\unit{u}) I_\nu(\unit{v}) \rangle - \langle I_\nu \rangle^2 \\ 
& \simeq & \int dz \,{{c\,dt} \over dz} \int dM {dn(z) \over dM} \,{1\over {\pi r_{sh}(M,z)^2}} \nonumber \\ 
& & \times \, \left[ {{\nu L_\nu(M,z)} \over {4 \pi (1+z)^4}} \right]^2 P_{1|2} \left[ {{\psi\,d_A(z)} \over {r_{sh}(M,z)}} \right] \coma \eeqrb
where $\unit{u}$ and $\unit{v}$ are unit vectors that satisfy $\unit{u} \cdot \unit{v} = \cos\psi$. The function $P_{1|2}$ is the probability that one line of sight passes through a halo of radios $r_{sh}$, given that another line of sight passes through the same halo (see \cite{Keshet04a} for detail). Halo-halo correlations, neglected in equation (\ref{eq:self_corr}), can only enhance the two-point correlation function. The \emph{fractional} correlation function, $\xi_\nu(\psi)\equiv \sqrt{\delta^2 I_\nu(\psi)} / \langle I_\nu \rangle$, is independent on the dynamical parameters $f_T$, $f_{acc}$, $\xi_e$ and $\xi_B$, and depends on the geometrical factor $f_{r}$ only through the relation\footnote{At low ($\nu<\nu_{br}$) or high ($\nu>\nu_{max}$) frequencies (see \S \ref{subsec:model_LCDM}), the scaling relation is more complicated.} $\xi_\nu(\psi;f_r) = f_{r}^{-1} \xi_\nu (f_r^{-1} \psi; 1)$.

\subsection{Fine-Tuned Model}
\label{subsec:model_LCDM}

The Press-Schechter mass function is known to disagree with numerical simulations, predicting less rare, massive halos and more abundant, low mass halos. The Press-Schechter approach agrees better with cosmological simulations, if the naive critical over-density $\delta_c(z)$ indicating collapse (calculated for spherical collapse) is replaced by \cite{Sheth2001}
\beq \widetilde{\delta_c}(z,M) = \sqrt{a}\,\delta_c(z) \left\{ 1 + b \left[ {\sigma^2(M) \over {a\, \delta_c(z)^2}} \right]^c \right\} \coma \eeq
where $\sigma(M)$ is the variance of the density field smoothed on a mass scale $M$, and $a$, $b$ and $c$ are parameters of order unity. Numerically, good agreement with \LCDM simulations is obtained if one chooses $a=0.73$, $b=0.34$, and $c=0.81$ \cite{Jenkins01,Barkana01}. Such a modification enhances the expected background from intergalactic shocks, because it increases the density of rare, massive halos, which dominate the extragalactic signal. We thus find a $20\%$ increase in the inverse Compton background, and a $50\%$ enhancement in the synchrotron signal. In what follows, we use the modified halo number density, unless otherwise stated. 

The spectrum of the integrated radiation from a halo accretion shock is essentially a broken power law, where the spectral break is introduced by inverse-Compton cooling. We focus on strong shocks which accelerate electrons to a power law energy distribution with index $p=2$, implying that the photon spectrum is flat ($\nu I_\nu \propto \nu^0$) at high frequencies and scales as $\nu I_\nu \propto \nu^{1/2}$ at low frequencies. 
The maximal Lorenz factor $\gamma_{max}$ to which electrons are accelerated implies maximal synchrotron frequencies around 
\beqr \label{eq:nu_max}
\nu_{max}(M,z) & \simeq & 3.6 \times 10^{14} \, \left[ {B(M,z) \over
{0.1\muG}} \right]^2 {T(M,z) \over {10^7\K}} \\ 
& & \times \,(1+z)^{-4} \Hz \fin \eeqr 
The spectral break frequency roughly corresponds to the minimal energy at which electrons manage to significantly cool (say, by $\eta=50\%$ of their initial energy) between the cosmic time of their host halo accretion shock (given by its redshift $z$) and the present epoch \cite{Keshet03}, 
\beqrb \label{eq:nu_min} \nu_{br}(M,z) & \simeq & 9 \, h_{70}^2 \left( {\eta \over 1-\eta} \right)^2 {B(M,z) \over {0.1\mu\mbox{G}}} \\ 
& & \times \left[ \int_0^z \frac{(1+z)^{3/2}}{g(a)}\,dz \right]^{-2} \kHz \fin \eeqrb 
The resulting synchrotron spectrum, integrated over all halos, cuts off below $\sim 100\MHz$. This implies that when including also weak shocks, the integrated spectrum $\nu I_\nu$ peaks at frequencies around $100 \MHz$.
  
A hydrodynamical \LCDM simulation \cite{Springel2001,Keshet03} was used to obtain rough estimates of the dimensionless free parameters used in the model \cite{Keshet04a}. For this purpose, various global features of the simulated \LCDM universe at the relevant epoch, $0<z<2$, were compared to their values according to the model. The calibration scheme was then tested by comparing features of the radiation from intergalactic shocks, as extracted from the simulation and as calculated from the model. The resulting parameter values are summarized in Table \ref{tab:ModelParams}. The energy fractions $\xi_e$ and $\xi_B$ can not be similarly evaluated from the simulation, but require independent observations (see \S\ref{subsec:model_efficiency}). 

The temperature parameter $f_T$ was estimated using the temperature statistics of the baryonic component of the universe, indicating that although $f_T$ has a weak redshift dependence, it lies in the range $0.45-0.55$ for the relevant epoch. The accretion rate parameter $f_{acc}$ was calibrated by the fraction of mass that has been processed by {\it strong} shocks in the relevant epoch. Comparing this fraction in the model and in the simulation indicated that $f_{acc}$ has a weak redshift dependence, and lies in the range $0.08-0.17$. The shock radius parameter $f_{r}$ has been estimated by studying the morphology of cluster accretion shocks, suggesting that $f_r$ lies roughly in the range $0.6-1.2$. 

An important aspect of the model is the sensitivity of the predicted synchrotron luminosity of a halo, but not its inverse-Compton luminosity, to halo asymmetry. Such asymmetries introduce spatial fluctuations in the thermal energy injection rate through the halo accretion shock, and enhance the synchrotron emission which scales as the \emph{square} of the thermal energy, $\nu L_\nu^{syn} \propto \dot{M} T^2$. This effect can be incorporated into the model by introducing a geometrical correction factor \cite{Keshet04a}. The localized nature of hot regions along the shock, and the large ratio $\zeta\simeq 10$ between the temperature of these regions and the temperature of the dimmer, more extended regions, justify approximating all the emission from a halo accretion shock as originating from an effective smaller region, of scale $\widetilde{r}_{sh} \simeq r_{sh}/\zeta$. Replacing $f_{r}$ with a different parameter, $\widetilde{f}_{r} \equiv f_{r}/ \zeta$, thus enhances the synchrotron luminosity of the halo by a factor $\zeta^2$, while leaving its inverse-Compton luminosity unchanged. As an order of magnitude estimate, $\widetilde{f}_{r}=f_{r}/\zeta \simeq 0.05-0.20$. Note that the two-point correlation functions are thus altered, power shifting to smaller angular scales.

\begin{table}[htb]
\caption{Intergalactic shock model parameters.}
\label{tab:ModelParams}
\begin{tabular}{@{}lll}
\hline
\em Par. & \em Quantity parameterized & \em Value (Range) \\
\hline
$f_T$     & Temperature & 0.50 (0.45-0.55) \\ 
$f_{acc}$ & Accretion rate  & 0.12 (0.08-0.17) \\ 
$f_{r}$  & Shock radius          & 0.9 (0.6-1.2) \\ 
$\widetilde{f}_{r}$ & Injection scale & 0.1 (0.05-0.2) \\
\hline 
$\xi_e$ & Electron energy fraction & 0.05 (0.02-0.10) \\
$\xi_B$ & Magnetic energy fraction & 0.01 (0.005-0.04) \\
\hline 
\end{tabular}\\[2pt]
Dynamical parameters are defined in \S \ref{subsec:model_review} and calibrated in \S \ref{subsec:model_LCDM}. 
Energy fractions are out of the shock thermal energy, and evaluated in \S \ref{subsec:model_efficiency}.
\end{table}

\begin{figure}[htb]
\begin{center}
\includegraphics[width=3in]{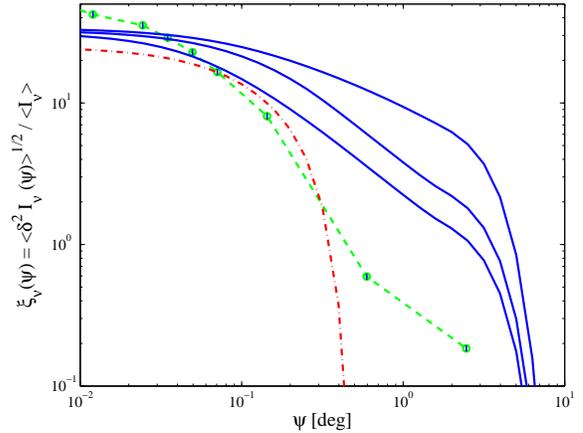}
\end{center}
\caption{ Fractional two-point correlation function $\xi_\nu(\psi,\nu)$ of synchrotron background from intergalactic shocks. The signal, calculated according to the calibrated model (see \S \ref{subsec:model_LCDM} and Table \ref{tab:ModelParams}) and shown for frequencies $1\MHz$, $100\MHz$ and $10\GHz$ (solid lines, bottom to top), scales roughly according to $\xi_\nu(\psi,\nu\,; \widetilde{f}_{r}) = \widetilde{f}_{r}^{-1} \xi_\nu (\widetilde{f}_{r}^{-1} \psi,\nu\,; 1)$ for $\nu_{br} \ll \nu \ll \nu_{max}$. The signal calculated from a \LCDM simulation (for $\nu=100\MHz$, neglecting emission below the break frequency $\nu_{br}$, circles with dashed line to guide the eye, error bars represent statistical errors introduced by averaging over random pairs of lines of sight \cite{Keshet04b}) is similar to the corresponding model result (dash-dotted line), yielding even more small scale power. 
}
\label{fig:calculated_correlation}
\end{figure}

With the $\Lambda$CDM-modified halo mass function and the best fit values described above, the model yields an inverse-Compton background $\nu I_\nu^{iC} \simeq 0.1 (\xi_e/0.05)\ICUnits$. The synchrotron spectrum is presented in Figure \ref{fig:LFRB_sky}, the two-point correlation function in Figure \ref{fig:calculated_correlation}, and the angular power spectrum (see \S\ref{subsec:feasibility_notations}) in Figure \ref{fig:synch_Cl}. It is encouraging to note that after calibrating the free parameters with essentially global features of the cosmological simulation, the model predictions agree well with results extracted from the simulation, regarding the \gama-ray background \cite{Keshet03}, the radio background, and the synchrotron two-point correlation function \cite{Keshet04b}. A more accurate calibration of the parameters, including an evaluation of their redshift dependence, may be obtained from a detailed analysis of the various clusters identified in cosmological simulations.

\subsection{Energy Conversion Efficiency}
\label{subsec:model_efficiency}

Intergalactic shock waves are expected to accelerate electrons to highly
relativistic energies, and to strongly amplify magnetic fields. The average
fractions of shock thermal energy deposited in relativistic electrons
($\xi_e$) and in magnetic fields ($\xi_B$) are both important parameters of
our model, each bearing linearly upon the predicted synchrotron signal, and
thus deserve a special discussion.

\subsubsection{Electron acceleration}
\label{subsec:model_electron_efficiency}

Collisionless, non-relativistic shocks are known to Fermi accelerate power-law energy distributions of relativistic particles. This phenomenon has been observed in astrophysical shock waves on various scales, such as in shocks forming when the supersonic solar wind collides with planetary magnetospheres, in shocks surrounding SNRs in the interstellar medium, and probably also in shocks in many of the most active extragalactic sources, quasars and radio galaxies \cite{Drury83,Blandford87}. The electron power-law distributions extend up to $\sim 100 \mbox{ TeV}$ energies in SNRs \cite{Tanimori98}, where shock velocities are of order $v \simeq 10^3 \km \se^{-1}$, similar to intergalactic shock velocities. 

Although no existing model credibly calculates the acceleration efficiency,
a simple argument \cite{Keshet03} suggests that we may evaluate $\xi_e$ in strong intergalactic shocks using the estimated acceleration efficiency of SNR shocks. Consider an ideally strong, non-relativistic shock wave, such that the shock Mach number $\Upsilon \gg 1$ and the thermal energy of the upstream plasma is negligible with respect to the shock energy. The physics of such a shock is essentially determined by three parameters: the shock velocity $v$, the upstream plasma number density $n_u$, and the upstream magnetic field strength $B_u$ (in principle, the result may also depend on the detailed structure of the upstream magnetic field). The upstream density may be eliminated from the problem altogether by measuring time in units of $\nu_{p}^{-1}$, where $\nu_{p}$ is the plasma frequency. 
The upstream magnetic field strength, parameterized by the cyclotron frequency, $\nu_{c}$, can not be additionally scaled out of the problem. However, comparing $\nu_{c,i}$ (where subscript $i$ denotes a property of the ions) to the growth rate of electromagnetic instabilities in the shocked plasma, $\nu_{ins} = \nu_{p,i} \,v /c$, indicates that their ratio in strong shocks satisfies $(\nu_{c,i}/\nu_{ins})^2 = (B_u^2/8\pi) / (n_u m_p v^2/2) \ll 1$. We thus \emph{assume} that there is a well behaved limit when this ratio approaches zero, implying that the upstream magnetic field has little effect on the behavior of strong shocks. With this assumption, we expect to find much similarity between sufficiently strong shocks in different environments, provided that their shock velocities are comparable, regardless of the plasma density and the strength or structure of the upstream magnetic field. The little effect of upstream magnetic fields on strong shocks is supported by recent SNR observations, which suggest that the shocks produce magnetic fields much stronger than found far upstream (see \S \ref{subsec:model_magnetic_efficiency}). 

We thus estimate the efficiency of strong intergalactic shocks from the strong SNR shocks, drawing upon the similarity between the velocities of the two families of shocks. The energy fraction deposited in relativistic electrons by SNR shocks was estimated by several authors, the most reliable estimates found for remnants with \gama-ray detection such as SNR1006 \cite{Tanimori98}. From such studies, we deduce an electron acceleration efficiency around $\xi_e \simeq 0.05$, with an uncertainty factor of $\sim 2$ (see \cite{Keshet04a} for details). An independent, less reliable method for estimating $\xi_e$ relies on the ratio between the energies of cosmic-ray electrons and cosmic-ray ions in the interstellar medium, suggesting that $\xi_e \simeq 1\%-2\%$ (for a discussion, see \cite{Keshet03}). However, since the relation between this ratio in the interstellar medium and immediately behind intergalactic (or SNR) shocks is unknown, this estimate is highly uncertain. 

We stress that our estimate of $\xi_e$ is based only on \emph{observations} of SNRs, without provoking any elaborate model for the acceleration mechanism of the electrons by the shocks. The high efficiency of electron acceleration suggests that a non-linear theory (e.g. \cite[and the references therein]{Berezhko99}) is required in order to account for the shock structure and the particle acceleration process. Non-linear theories for diffusive acceleration of particles by shock waves are at the present stage incomplete (for a recent review, see \cite{Malkov01}). In particular, such models fail to produce the flat electron spectra, which are implied by the observations and predicted by a linear, test-particle approximation.

\subsubsection{Magnetic field amplification}
\label{subsec:model_magnetic_efficiency}

Magnetic fields in galaxy cluster halos have been estimated using several techniques, based on diffuse synchrotron emission from the cluster, preferably combined with inverse-Compton detection, on Faraday rotation of background or embedded polarized radio sources, and on the observation of cold fronts in cluster X-ray images (for reviews, see \cite{Kronberg94,Henriksen98,Carilli02}). The synchrotron emission from a cluster measures the volume-averaged magnetic field weighted by the relativistic electron distribution, and is thus most relevant for our model. Studies of the emission suggest volume averaged magnetic field strengths of the order of a few $0.1\muG$, close to the values inferred for a minimal energy configuration with equal energy in magnetic fields and in relativistic electrons. Faraday rotation and cold front studies suggest stronger magnetic fields of several $\mu$G, but are sensitive to different measures of the magnetic field and to the assumptions made (see \cite{Keshet04a} for discussion). 

A fraction $\xi_B\simeq0.01$ of a shock thermal energy transferred to magnetic energy, implies magnetic field strengths that are an order of magnitude lower than their equipartition value (with respect to the thermal electrons, hereafter), and are close to their value in a configuration with equal energy in magnetic fields and in relativistic electrons (assuming $\xi_e \simeq 0.05$). In cluster halos, our calibrated model reproduces, for this choice of $\xi_B$, a volume-averaged magnetic field strength $B\simeq 0.1\muG$ (for $M\simeq 10^{14}M_\odot$), although the 'bright spots' could contain magnetic field strengths as high as $1\muG$. The lowest observational estimates for cluster magnetic field strengths are $>0.05\muG$, suggesting that $\xi_B$ could be higher than our chosen value by a factor of a few, but is unlikely to be smaller than it by more than a factor of $\sim 2$.

It is important to note that recent studies suggest that SNRs contain
strong, near equipartition magnetic fields. For example, the highly localized
nature of hard X-ray emission in the resolved sheets composing the shock front
of SNR1006 strongly suggests the presence of strong, near equipartition magnetic fields behind the shock (e.g. \cite{Berezhko03}). The narrow extent of emission observed upstream of these sheets implies a small upstream diffusion constant, corresponding to upstream magnetic fields of strength $\mga 10\muG$, with strong fluctuations ($\delta  B/B\simeq 1$) on small, $d \ll 10^{17}\cm$ scales \cite{Bamba03}. Such magnetic field fluctuations are far stronger, and have much smaller scales, than found in the surrounding ISM, and so must be induced by the shock. These conclusions strongly suggest that the assumption, that the physics of a strong shock depends only weakly upon the far upstream magnetic field, is valid, justifying the analogy between strong intergalactic shocks and strong SNR shocks of comparable velocities.

\subsection{Radio Halo Observations}
\label{subsec:model_halos}

\begin{figure}[htb]
\begin{center}
\includegraphics[width=3in]{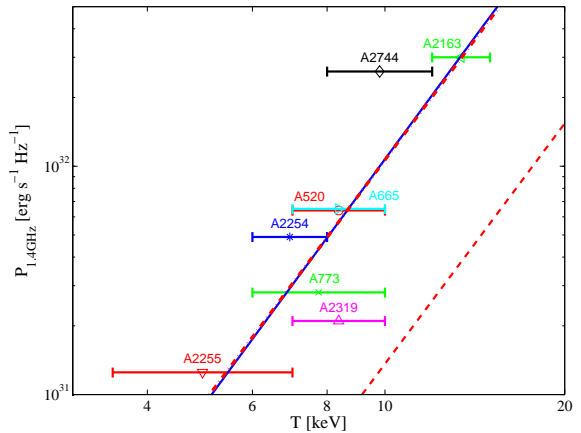}
\end{center}
\caption{ Specific luminosities of radio halos for $\nu = 1.4\GHz$, plotted against the temperature ranges of their host galaxy clusters (error bars with cluster names, adopted from \cite{Govoni01,Feretti03}). The best power law fit to the data (solid line) scales as $L_\nu (T,\nu = 1.4 \GHz) \propto T^{3.55}$. The specific luminosity according to the model scales similarly, $L_{\nu}(T) \propto T^{7/2}$, but the normalization (of the calibrated model, dashed line) is different; the dash-dotted line presents the result of the model where the temperature was re-scaled by a factor of $1.8$. 
}
\label{fig:clusters}
\end{figure}

Radio halos are observed in $\sim 35\%$ of the young, massive galaxy clusters (with X-ray luminosity $L_X>10^{45}\erg\se^{-1}$, see \cite{Giovannini99,Feretti03}), and their radio luminosity is known to be correlated with the cluster temperature \cite{Liang01} and mass \cite{Govoni01}. Figure \ref{fig:clusters} shows the specific luminosities of such halos for $\nu = 1.4 \GHz$, plotted against the temperature ranges of their host clusters. A simple power law fit to the data, $L_\nu(T, \nu=1.4\GHz) \propto T^{\phi}$, gives $\phi=3.55$, which is very similar to the power law index $\phi=7/2$ predicted by our model, although the (calibrated) model yields a specific luminosity that is lower than observed by a factor of $\sim 8$ (equivalently, the temperatures are discrepant by a factor of $\sim 1.8$). A similarly good agreement between the model and the data is obtained when replacing the cluster temperatures with their estimated masses, giving $L_\nu (M, \nu=1.4 \GHz) \propto M^{2.2}$ \cite{Govoni01}, similar to the model prediction $L_\nu^{syn} (M) \propto M^{7/3}$. However, the estimated mass of a cluster is somewhat less certain than the measured temperature, and is sensitive to the definition of the cluster boundary.

Some comments regarding Figure \ref{fig:clusters} are in place. First note, that the agreement between model and observations regarding the value of $\phi$ is insensitive to the free parameters of the model. The specific synchrotron luminosity of the model, when written as a function of halo temperature, scales linearly with the combination $f_{acc} f_{r}^{-2} f_T^{-3/2} \xi_e \xi_B$. The strong dependence of this result upon the model parameters, and the low redshifts ($z \mla 0.3$) of all the halos shown in Figure \ref{fig:clusters}, suggest that the discrepancy between model and observations may be eliminated by better calibrating the parameters, and modelling their redshift dependence (note that this will \emph{increase} the radio signal we predict). Second, note that the low fraction of clusters with observed radio halos could result from a combination of two effects: (i) not all clusters have been significantly accreting new mass at the cosmic time when they are observed; and (ii) the surface brightness of the halos observed is low, close to the instrumental sensitivity, suggesting that more radio halos will be observed with future telescopes. A varying accretion rate will tend to enhance the luminosity of the halos that are observed, possibly accounting for some of the discrepancy seen in Figure \ref{fig:clusters}. Finally, note that in \cite{Minitai01}, a cosmological simulation was used to show that $\phi=2.6-2.8$ (for emission from electrons accelerated at shocks; for emission from secondary electrons produced by $p$-$p$ collisions of cosmic ray ions, they found $\phi=4.1-4.2$). However, the simulation of \cite{Minitai01} differs substantially from our model and simulation \cite{Keshet03,Keshet04b}, regarding their cosmological model (SCDM), electron acceleration efficiency ($\xi_e< 0.5\%$, see also \cite{Miniati02}), and magnetic field normalization ($\langle B^2 \rangle^{1/2} \simeq 3\muG$ for a Coma-like cluster).


\section{Observational Consequences} 
\label{sec:feasibility}
\indent

In this section we examine the capability of present and next generation radio telescopes, in particular the SKA, to detect the radio emission from intergalactic shocks. In \S \ref{subsec:feasibility_notations} we present the notations used. Various foreground and background signals are discussed in \S \ref{subsec:feasibility_signals}, in an attempt to assess the optimal
conditions for detecting the emission from intergalactic shocks.

\subsection{Notations}
\label{subsec:feasibility_notations}
\indent 

As mentioned in \S \ref{sec:IGM_model}, the two-point correlation function of specific intensity fluctuations at a given angular separation $\psi$, is defined as 
\beqr \delta^2 I_\nu (\psi) & \equiv & \langle \delta I_\nu(\unit{u}) \delta I_\nu(\unit{v}) \rangle \\ & = & \langle I_\nu(\unit{u}) I_\nu(\unit{v}) \rangle - \langle I_\nu \rangle^2 \coma \eeqr
where $\unit{u}$ and $\unit{v}$ are unit vectors that satisfy $\unit{u} \cdot \unit{v} = \cos\psi$, and $\delta I_\nu (\unit{u}) \equiv I_\nu(\unit{u}) - \langle I_\nu\rangle$. The specific intensity is related to the brightness temperature $T_b$ (which approximately equals the thermodynamic temperature for the sky brightness and the frequency range of interest) by $I_\nu = 2 \nu^2 k_B T_b /c^2$. The following discussion of intensity fluctuations is thus equally applicable for temperature fluctuations, up to a multiplicative constant. 

\begin{figure}[htb]
\begin{center}
\includegraphics[width=3in]{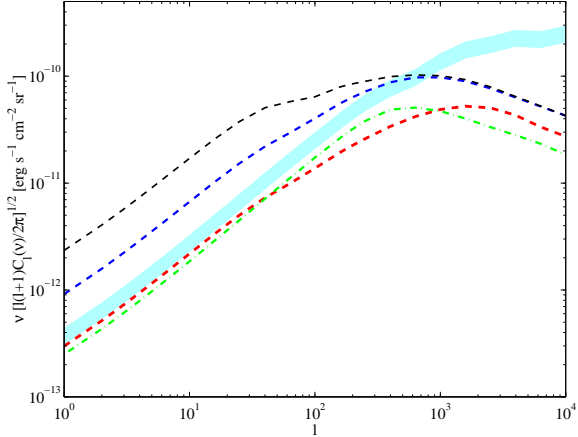}
\end{center}
\caption{ Angular power spectrum of the synchrotron background from intergalactic shocks, calculated according to the calibrated model (see \S \ref{subsec:model_LCDM} and Table \ref{tab:ModelParams}) for frequencies $1\MHz$, $100\MHz$, and $10\GHz$ (dashed lines, bottom to top). The result of a \LCDM simulation for $\nu=100\MHz$ (neglecting emission below the spectral break \cite{Keshet04b}, shown as a band with thickness corresponding to the statistical error) exhibits more power than the corresponding result of the model (dash dotted line), in particular on small, arcminute scales.  }
\label{fig:synch_Cl}
\end{figure}

It is often advantageous to study the \emph{Angular Power Spectrum} (hereafter APS) of various signals, in order to obtain a direct estimate of their importance at various angular scales. The angular power spectrum $C_l$ is defined through the relation 
\beq \label{eq:corr_func} \delta^2 I_\nu (\psi) \equiv \frac{1}{4\pi} \sum_{l=1}^{\infty} (2l+1)C_l(\nu) \,P_l(\cos\psi) \coma \eeq
where $P_l(x)$ is the Legendre polynomial of degree $l$. The power at a given multipole is often expressed using its logarithmic contribution to the intensity variance, $\delta I_l\equiv [l(2l+1)C_l/4\pi]^{1/2}$ (or $\delta T_l$ for the variance of the brightness temperature), where multipole $l$ roughly corresponds to angular scales $\theta \simeq 180\de/l$. Using the orthogonality of the Legendre polynomials, we may invert equation (\ref{eq:corr_func}) to find the APS, 
\beq \label{eq:Cl_from_xi} C_l(\nu) = 2\pi \int_0^\pi \delta^2 I_\nu(\psi) P_l(\cos \psi) \sin \psi \,d \psi \fin \eeq
The APS of synchrotron emission from intergalactic shocks may be calculated from the two-point correlation function (see Figure \ref{fig:calculated_correlation}) using equation (\ref{eq:Cl_from_xi}). The resulting APS, presented in Figure \ref{fig:synch_Cl}, peaks at multipoles $l\simeq 400-4000$ for the relevant frequency range, corresponding to angles $\theta \simeq 3^\prime-30^\prime$.

The sensitivity of a telescope is often estimated using its RMS noise level for a given beam size $\sigma = \pi \psi^2$, denoted $\delta I_{rms} \equiv [\delta^2 I(\sigma)]^{1/2}$. For well behaved signals, where the two-point correlation function $\delta^2 I(\psi)$ is a monotonically decreasing function of $\psi$, one finds $\delta^2 I(\psi)<\delta^2 I_{rms}(\sigma)$. Hence, the RMS noise level imposes an upper limit to the two-point correlation function sensitivity, and may be compared directly with the logarithmic contribution to the variance at the corresponding multipole, $\delta I_l$.

\subsection{Competing Signals}
\label{subsec:feasibility_signals}
\indent

In the following, we examine various foreground and background signals in the low frequency radio band. We focus on Galactic synchrotron foreground and on discrete radio sources, which contaminate the expected fluctuating signal on large and on small angular scales, respectively. Other extragalactic low frequency signals --- bremsstrahlung from \Lya clouds and 21 cm tomography --- are also discussed. The contribution of each signal to the sky brightness is shown in Figure \ref{fig:LFRB_sky}, and its contribution to the variance on $0\de.5$ angular scales in Figure \ref{fig:Sky_Cl}.

\subsubsection{Galactic Synchrotron Emission} 
\indent

At low frequencies, $\nu \mla 1\GHz$, the brightness of the sky is dominated
by Galactic synchrotron emission, produced by cosmic-ray electrons gyrating
in the magnetic fields of the interstellar medium (ISM). At low Galactic latitudes, there is also a substantial contribution of free-free emission from low-latitude \HII regions (e.g \cite{Baccigalupi01}). In the frequency
range of interest, $\nu\ll10\GHz$, the two-point correlation function is
dominated by Galactic synchrotron emission, because of its significant
power on large angular scales (see \cite{Keshet04a} for discussion). We thus focus on the contamination by Galactic synchrotron emission in high Galactic latitudes and on small angular scales.

With large uncertainties regarding the spatial distribution of Galactic
cosmic rays and magnetic fields, the best estimates of Galactic synchrotron
emission are obtained by extrapolating direct measurements carried out at
frequencies and angular scales where synchrotron emission is reliably
measured. We thus make the key assumption, often used in the literature, that the multipole dependence of the Galactic synchrotron APS varies little with frequency, such that $C_l(\nu) \simeq f(l) g(\nu)$. The frequency dependence $g(\nu)$ of synchrotron emission can be modelled as a broken power law, and is discussed in \cite{Keshet04a}. The APS multipole dependence $f(l)$ may be extracted from high resolution maps at $\sim\GHz$ frequencies down to arc-minute scales (multipoles $l\simeq 6000$, \cite{Tucci02}), and may in principle be extrapolated into the entire frequency regime and multipole range of interest.

The angular power spectra extracted from radio maps depend on a number of factors, most importantly the Galactic latitude range examined and the efficiency at which discrete radio sources are removed from the map. Unfortunately, high resolution surveys are available mostly for low latitudes, which are less promising as potential search sites for an extragalactic signal because they exhibit a stronger Galactic foreground, in particular at small angular scales. Nonetheless, improved high and medium latitude data have recently made it possible to extend our knowledge of the high latitude APS up to multipoles $l\simeq 800$. 

The high latitude ($60.5\de<b<84.5\de$) low resolution (FWHM $0.6\de-2.3\de$) maps of \cite{Brouw76} for 5 frequencies between $408\MHz$ and $1411\MHz$, have been used to estimate the APS up to $l\simeq 70$ (e.g. \cite[table 6]{Bruscoli02}). Analysis of the Bonn $408\MHz$ full-sky survey \cite{Haslam82} suggests a power-law APS of the form $C_l(l\gg 1)\propto l^{-\beta}$ with $\beta \simeq 2.5-3.0$ down to the survey resolution limit $0.85\de$, corresponding to multipoles $l\simeq 200$ \cite{Tegmark96}. With a characteristic coherence scale $\sim 5\de-10\de$ for synchrotron emission (e.g. \cite{Spoelstra84,Banday91}), $C_l\propto (l+5)^{-3}$ was claimed \cite{Tegmark96} to describe the APS rather well. Thus, although the contribution of Galactic synchrotron emission to the two-point correlation function is significant, it is dominated by large angular scales $\sim 5\de-10\de$ (low multipoles $l < 40$) and introduces \emph{little} contamination at small angular scales (see Figure \ref{fig:Sky_Cl}). The APS exhibits strong fluctuations across the sky, suggesting that 'quiet' regions may be identified and selected as preferable search sites for an extragalactic signal.  

An analysis of the $2.3 \GHz$ Rhodes map \cite{Jonas98}, with FWHM
resolution $20^\prime$, reveals strong variations of the APS with Galactic
latitude, and a large contribution from discrete radio sources which tend
to flatten the APS \cite{Giardino01}. At high latitudes ($|b|>20\de$), removal of point sources using median filtering gives \cite{Giardino01} 
\beq C_l \simeq A^2 l^{-\beta} \coma \eeq with $A=0.3\pm 0.2\K$, and $\beta=2.92\pm0.07$, valid up to $l\simeq 100$. Although the quoted uncertainties are large, comparison with the low resolution studies
mentioned above indicates that the normalization $A$ is probably no larger
than $0.3\K$. Similarly steep APS, with $\beta$ varying in the range
$2.60-3.35$ for different medium latitude ($|b|\mla 20\de$) regions, were
found from a $1.4\GHz$ survey carried out with the Effelsberg $100$ m
telescope \cite{Uyaniker99}, after removing $>5\sigma$ unresolved point
sources \cite{Baccigalupi01}. The high angular resolution of the Effelsberg
survey, $\sim 9.35^\prime$, thus suggests that the steep power law
discussed above holds up to multipoles $l\simeq 800$. Note that after removing the brightest intensity peaks associated with low latitude \HII regions, the APS highly resembles the APS of the polarized intensity component \cite{Baccigalupi01}, as expected for synchrotron emission.

\subsubsection{Discrete Radio Sources}

Discrete radio sources, mostly radio galaxies, active galactic nuclei
(AGNs) and normal galaxies, make an important contribution to the
extragalactic radio background. Catalogues of radio sources at several
frequencies have been used to estimate their contribution to the radio sky
\cite{Simon77,Willis77}. This procedure is limited by the uncertain
contribution of faint, unidentified sources (mostly normal galaxies), which
dominate the background (see for example \cite{Windhorst93}) and must be
modelled. Other studies have estimated the contribution of discrete sources
to the radio background, utilizing the well-known correlation between the
infra-red and radio flux densities of individual galaxies
\cite{Protheroe96,Haarsma98}. However, the results of such studies are
sensitive to the unknown redshift evolution of the sources. In addition,
the radio-infra-red correlation holds only in cases where the radio
emission is associated with star formation. Hence, such studies must be
supplemented by independent estimates of sources not associated with star
formation, such as a catalogue-based estimate of the emission from AGNs
\cite{Ryle68}.

\begin{figure*}[p]
\begin{center}
\includegraphics[width=6in]{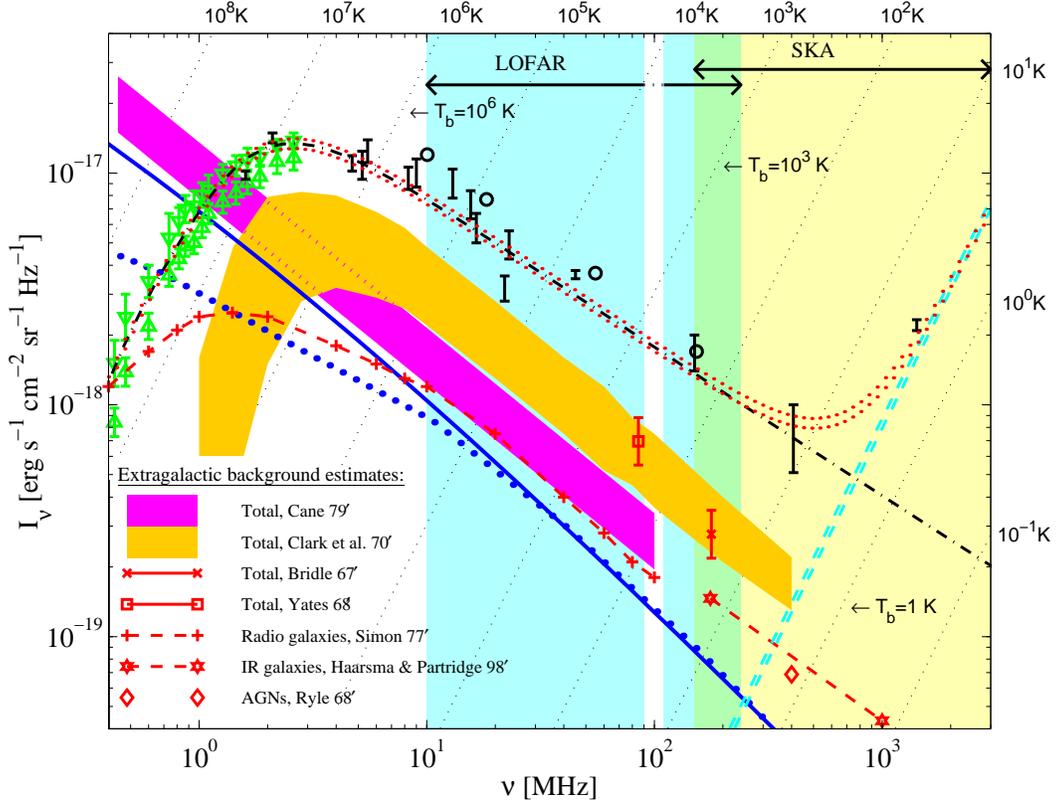}
\end{center}
\caption{ The low frequency radio sky. Ground-based (circles and error bars for the Galactic polar regions) and space-based (triangles for IMP-6 spacecraft minimum and maximum) observations are dominated by Galactic synchrotron emission (dash dotted line shows the modelled Galactic foreground toward the polar regions \cite{Keshet04a}) in low frequencies, and by the CMB (double-dashed line) at high frequencies. Various estimates of the extragalactic component (legend, see text) are typically an order of magnitude lower than the Galactic foreground. The background from intergalactic shocks according to the calibrated model (of \S \ref{subsec:model_LCDM} and Table \ref{tab:ModelParams}, solid line) is roughly of the same magnitude, scaling according to $I_\nu\propto f_{acc} f_T^2 \widetilde{f}_r^{-2}$. This signal is also shown according to a \LCDM simulation (accounting only for emission above the spectral break, dotted line, \cite{Keshet04b}). The frequency ranges of the SKA and the LOFAR (double arrows and corresponding shaded regions) and brightness temperature contours (dotted lines, labelled on the top and on the right axes) are also shown.  }
\label{fig:LFRB_sky}
\end{figure*}

The studies outlined above suggest that the contribution of discrete sources to the radio sky is roughly an order of magnitude lower than the observed sky brightness. This corresponds to $25\%-65\%$ of the total extragalactic background, as estimated by spectral modelling of low frequency radio observations \cite{Clark70,Cane79,Keshet04a}, although some evolutionary models (e.g. \cite{Protheroe96}) may even account for the entire strong extragalactic signal calculated by \cite{Clark70}. The spectral index of the integrated emission from point sources has been estimated to lie in the range $s=0.7-0.8$ for frequencies in the range $100\MHz-\mbox{few} \GHz$ \cite{Simon77,Lawson87,Haarsma98}, although some dependence upon scale may be expected because the source distribution is essentially bimodal \cite{Tegmark00}. At very low ($\nu\mla10\MHz$) frequencies, the background from discrete sources turns around, probably because of synchrotron self-absorption \cite{Simon77}. 

It is easier to estimate the contribution of point sources to the APS,
because the latter is dominated by the bright, well studied
sources. Discrete radio sources have approximately a Poisson distribution
in the sky, because projection through their wide redshift distribution
effectively diminishes their correlations (e.g. \cite{Tegmark96}). Hence, as long as the angular scales concerned are much larger
than the angular extent of the sources, the two-point correlation function
vanishes and the APS is flat, $C_l\propto l^0$ (white noise), such that
roughly $\delta I_l\propto l$. One may estimate the APS as (e.g. \cite{Tegmark96})
\beq \label{eq:Cl_Poisson} C_l = \int_0^\infty S^2 {\partial N \over
\partial S} \,dS \coma \eeq where $S$ is the source flux and $\partial
N/\partial S$ is the differential number density of point sources in the
sky. The limited sensitivity $S_{min}$ of the observations limits our
knowledge of $\partial N/\partial S$ to sources with flux $S>S_{min}$, but
since the number density of faint sources is not too steep, the possible
error introduced is small. The upper limit of the integral, essentially
determined by the brightest sources, can be lowered in order to reduce the
noise, by modelling and removing the brightest sources from the analyzed
map. However, this quickly becomes laborious as the number of sources
increases, and introduces inevitable errors associated with source removal
uncertainties.

\begin{figure*}[p]
\begin{center}
\includegraphics[width=5.5in]{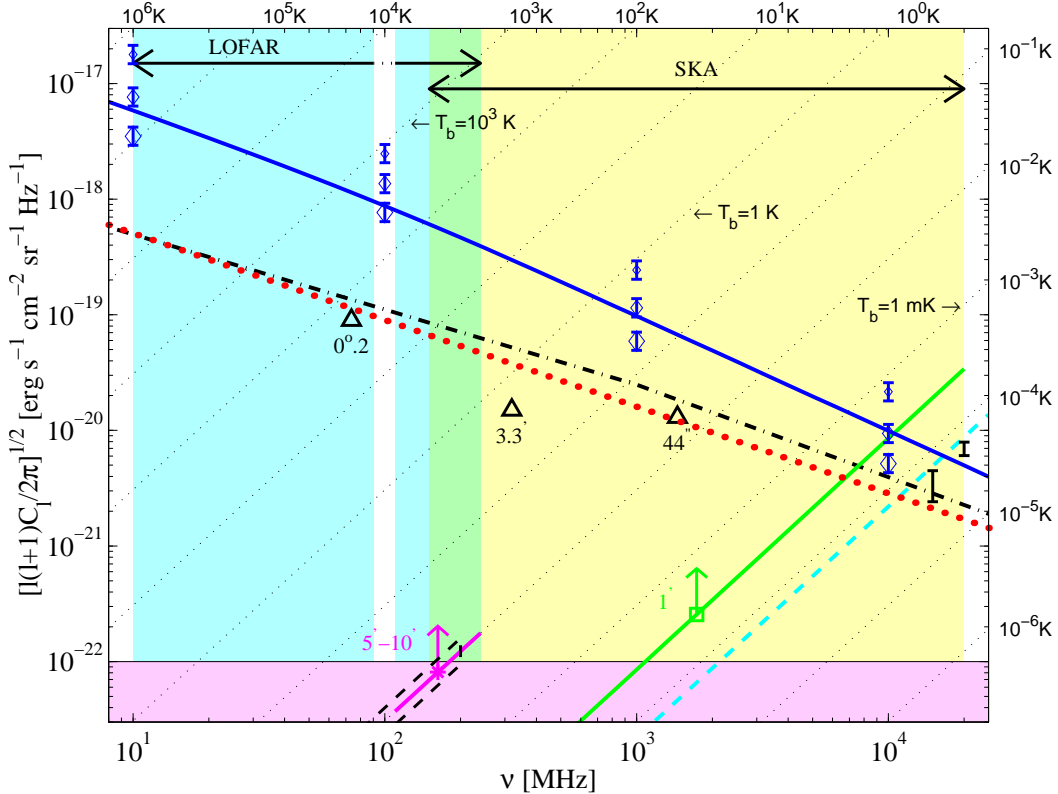}
\end{center}
\caption{ Logarithmic contribution $\delta I_l$ of various signals to the variance at multipole $l=400$ (corresponding to $\theta\simeq 0\de.5$). The background from intergalactic shocks is shown according to the calibrated model of \S \ref{subsec:model_LCDM} and Table \ref{tab:ModelParams} (solid line), and scales roughly according to $\delta I_l(l; \, \xi_e , \xi_B , f_{acc},f_T,\widetilde{f}_{r}) = \xi_e \xi_B f_{acc} f_T^2 \widetilde{f}_{r}^{-3} \,\delta I_l(\widetilde{f}_{r}l;\,1,1,1,1,1)$. The signal according to a \LCDM simulation (diamonds with statistical error bars for $l=400$, $10^3$ and $10^4$, from bottom to top, accounting only for emission above the spectral break, see \cite{Keshet04b}) scales linearly with the combination $\xi_e \,\xi_B$. The high-latitude Galactic foreground (dash-dotted line) and the integrated emission from discrete radio sources (dotted line, assuming removal of sources above $S_{cut}=100\mJy$ at $1.4\GHz$) are presented. Above $\sim10\GHz$, CMB fluctuations take over (dashed line extrapolated from WMAP measurements, error bars present detections by the Cambridge Radio Telescope on $0\de.2-0\de.5$ scales, and by the Owens Valley Radio Observatory on $0\de.1-0\de.6$ scales; for references see \cite{Hinshaw03,White99}). Bremsstrahlung emission from \Lya clouds (low horizontal shaded region) and the maximal 21 cm tomography signal (dashed contour) are seen to be much weaker than the above competing signals. Continuum surface brightness sensitivities of the SKA (square) and the LOFAR (star), and the side-lobe confusion limit of the VLA (triangles, for 10 minutes in configuration D, see http://www.vla.nrao.edu/astro/guides/vlas) are shown (labels denote beam widths). Brightness temperature contours are shown as dotted lines (labelled on the top and the right axes). Emission from intergalactic shocks is seen to dominate the sky at this angular scale for frequencies $\nu \mla 500\MHz$, provided that $\xi_e \, \xi_B \mga 10^{-4}$. According to the simulation, the emission dominates the sky at this frequency range for $\theta <0\de.5$, even if $\xi_e \, \xi_B = 5\times 10^{-5}$.  }
\label{fig:Sky_Cl}
\end{figure*}

Following \cite{Tegmark96}, we use source counts produced by the $1.4\GHz$ VLA FIRST all-sky survey. A small fraction ($\sim 10^{-4}$) of the discrete sources in the FIRST catalogue have angular scales larger than an arc-minute \cite{White97}, so the use of equation
(\ref{eq:Cl_Poisson}) is legitimate when dealing with multipoles $l\mla
10^4$. The number density of radio sources (after performing a
resolution correction for extended sources, \cite{White97}) is well fit in the
flux range $S=1\mJy-1\Jy$ by 
\beqr {\partial N \over \partial S} & \simeq & 1.5\times 10^8 \left(S \over 1\mJy \right)^{-1.5} \\ 
& & \times \left(1 + {S \over 100 \mJy} \right)^{-1} \Jy^{-1} \sr^{-1} \fin \eeqr 
For faint sources in the range $S=10\muJy-1\mJy$, the number count steepens to $\partial N /\partial S \propto S^{-\gamma}$ with $\gamma = 2.2\pm 0.2$ \cite{Windhorst93}. This changes the integrated brightness considerably, but has a negligible effect on the APS. Using equation (\ref{eq:Cl_Poisson}), we find that for a modest cut at $S_{max}=100\mJy$, corresponding to removal of the $\sim 7\times 10^4$ brightest sources all-sky, $C_l(\nu) \simeq 1.7 \times
10^{-8} (\nu/1.4\GHz)^{-2.75} \K^2$. Using the 6C survey carried out at $151\MHz$ \cite{Hales88}, we find that the above cut is equivalent to
removal of sources brighter than $\sim 3\Jy$ at $150\MHz$. As shown in Figure
\ref{fig:Sky_Cl}, such a cut places the emission from point sources at the
same level as the Galactic foreground for angular scales $\sim 0\de.5$. At
smaller angular scales, contamination from discrete sources becomes
increasingly worse, requiring the removal of more sources.

\subsubsection{Bremsstrahlung from \Lya Clouds}
\label{sec:brehmsstrahlung}
\indent

The thermal bremsstrahlung emission from \Lya clouds has been calculated in \cite{Loeb96}. These clouds are most likely in photo-ionization equilibrium with the UV background, permitting an estimate of the bremsstrahlung signal if the UV background intensity is known. UV background intensity estimates based on the proximity effect \cite{Scott02} imply a bremsstrahlung signal \cite{Keshet04a}
\beq \label{eq:total_brem_flux} J_\nu = 10^{-22 \pm 1} \IUnits \coma \eeq
for $\nu \ll 10^{14} \Hz$. This result neglects the contribution of optically thick clouds at high ($z>5$) redshifts, and possible additional ionization fields, such that the actual bremsstrahlung flux may be somewhat higher than given above. 

The amplitude of intensity fluctuations introduced by the finite number of \HI clouds along any given line of sight has also been evaluated \cite{Loeb96}. Assuming a random cloud distribution in redshift and in column density, recent UV background estimates imply RMS fluctuations \cite{Keshet04a}
\beq \Delta J_\nu = 10^{-23\pm1} \IUnits \coma \eeq 
for $\nu \ll 10^{14} \Hz$. Such fluctuations are expected on $\sim 7^{\prime \prime}$ scales (corresponding to $l\simeq 10^5$), although weaker fluctuations introduced by clouds associated with large-scale structure may appear even on $\sim 10^\prime$ scales. This fluctuation level should be regarded as a lower limit, because in addition to the assumptions leading to equation (\ref{eq:total_brem_flux}), the non-randomness of the cloud distribution further enhances the fluctuation signal. Nevertheless, without a unique spectral or temporal signature, confusion with the other signals discussed in this section, in particular discrete sources, is likely to preclude detection of the bremsstrahlung signal in the relevant frequency range.

\subsubsection{IGM 21 cm Tomography}
\label{sec:21tomography}
\indent

The 21 cm ($1.4\GHz$) spin flip transition of atomic hydrogen was proposed as  a probe of the IGM at the epoch prior to reionization \cite{Madau97} (see also \cite{Tozzi00}) . Spatial inhomogeneities in the IGM may be observed today as redshifted emission or absorption fluctuations against the CMB, in situations where early sources of radiation decoupled the IGM spin temperature from the CMB temperature. In principle, the combined angular and spectral signal, stronger than the CMB fluctuations by two orders of magnitude, may be used to trace the 'cosmic-web' structure of the early universe, in both space and cosmic time. 

However, as pointed out by \cite{DiMatteo02}, at the relevant frequency range $50-200\MHz$, contamination by radio point sources imposes a serious contamination even for the maximal signal amplitude $\Delta
T\simeq 10\mK$ expected. A \emph{spectral} $\sim10\mK$ feature, caused by the fast rise of the \Lya background as the first UV sources reionized the IGM, could possibly be detected behind the spectrally continuous foreground \cite[and references therein]{Zaldarriag03,Morales59,Loeb2003,Gnedin03}. Emission from intergalactic shocks provides an additional important source of confusion for $21\cm$ tomography.


\section{Discussion}
\label{sec:discussion}

We have calculated the imprint of intergalactic shocks on the low frequency radio sky, using a simple model, calibrated and tested with a \LCDM simulation. Our main results are illustrated in Figures \ref{fig:LFRB_sky} and \ref{fig:Sky_Cl} (for the calibrated model parameters summarized in Table \ref{tab:ModelParams}). Figure \ref{fig:LFRB_sky} shows the contribution of various signals to the low frequency radio sky, suggesting that emission from intergalactic shocks contributes up to a few tens of percent of the extragalactic radio background below $500\MHz$. Figure \ref{fig:Sky_Cl} depicts the angular power spectrum of various signals on an angular scale of $\sim0\de.5$, along with the sensitivity and the angular resolution of the SKA. 

We conclude that the design of the SKA, as well as the design of next generation radio telescopes such as the LOw Frequency Array (LOFAR\footnote{see http://www.lofar.org}) and the Astronomical Low Frequency Array (ALFA\footnote{see http://sgra.jpl.nasa.gov/html\_dj/ALFA.html}), are more than sufficient for detection of the angular fluctuations introduced by intergalactic shocks, as calculated in \S \ref{sec:IGM_model} (see also \cite{Keshet04a,Keshet04b}). Identification of the signal is limited by confusion with Galactic foreground and with discrete radio sources. Foreground
fluctuations in the synchrotron emission of our Galaxy constrain a clear
detection of the signal to sub-degree scales and to high Galactic 
latitudes. Confusion with discrete radio sources requires that the
brightest sources be modelled and removed. With a feasible point source cut
($100\mJy$ at $1.4 \GHz$, or $3\Jy$ at $150\MHz$), the signal dominates
over the competing signals at angular scales $10\marcmin \mla \theta \mla
1\de$ (in low frequencies, for $\xi_e\,\xi_B \mga 5\times 10^{-5}$, see discussion below), whereas detection on arcminute scales will require a more ambitious point source removal. The spectrum of intergalactic shock emission indicates that the signal is most pronounced at frequencies around $\sim 100\MHz$, planned to be covered by the SKA, the LOFAR and the ALFA. 

The calculated signal is sensitive to uncertainties in the model parameter calibration. The logarithmic contribution to the variance (shown
in Figure \ref{fig:Sky_Cl}) scales roughly according to 
\beqrb \delta I_l (l;\,\xi_e,\!\!\!\!&\xi_B&\!\!\!\!\!,f_{acc},f_T,\widetilde{f}_{r}) \\ & = & \xi_e \xi_B {f_{acc} f_T^2 \over \widetilde{f}_{r}^3} \delta I_l (\widetilde{f}_{r} l; \,1,1,1,1,1) \fin \eeqrb 
The radio signal calculated from the cosmological simulation \cite{Keshet04b}, however, depends only on the energy fractions $\xi_e$ and $\xi_B$, scaling linearly with the combination $\xi_e \,\xi_B$. Comparison between the predictions of the model and of the simulation (e.g. in Figures \ref{fig:calculated_correlation}, \ref{fig:synch_Cl} - \ref{fig:Sky_Cl}) suggests that $C_l$ was in fact \emph{underestimated} by our parameter calibration scheme. This implies, for example, that intergalactic shocks introduce intensity fluctuations of magnitude $\delta I_l \mga 8\times 10^{-19} (\xi_e \, \xi_B / 5\times 10^{-4}) (\nu/100\MHz)^{-1} \IUnits\!\!$ on multipoles $400\mla l \mla 2000$, corresponding to temperature fluctuations $\delta T_l \mga 260 (\xi_e \, \xi_B / 5\times 10^{-4}) (\nu/100\MHz)^{-3} \K$. 

The fluctuations introduced by intergalactic shocks, as estimated above, dominate the sky at $<500\MHz$ frequencies if $\xi_e \,\xi_B\mga 10^{-4}$, and are dominant even at frequencies as high as $10\GHz$ if $\xi_e \,\xi_B\mga 3 \times 10^{-4}$. The simulation suggests that the signal is stronger than predicted by the model, in particular on small angular scales. Therefore, according to the simulation, emission from intergalactic shocks will dominate the sky at $<0\de.5$ scales and $<500\MHz$ frequencies, even if $\xi_e \,\xi_B \simeq 5\times 10^{-5}$. We have used observations of SNR shocks and of magnetic fields in the halos of galaxy clusters, in order to show that for strong intergalactic shocks $\xi_e \,\xi_B \simeq 5 \times 10^{-4}$, unlikely to be smaller than this value by more than a factor of $\sim 4$ (see \S \ref{subsec:model_efficiency}). Detection of the signal is not unrealistic even in case $\delta I_l$ has been mildly overestimated. A signal lower than calculated in \S\ref{sec:IGM_model} by a factor of a few, may still be identified at $\sim 1\marcmin-10\marcmin$ scales, if faint point sources are modelled and removed. If the signal was overestimated by $\sim 2$ orders of magnitude, it may still be detectable by next generation radio telescopes such as the SKA, by means of cross-correlation with known tracers of large-scale structure, such as galaxy counts, and in the future, with \gama-ray emission from intergalactic shocks. 

Interestingly, detection of the intergalactic shock signal is just possible with \emph{present} high resolution radio telescopes, such as the VLA (at the maximal sensitivity configuration, see Figure \ref{fig:Sky_Cl}). The calculated signal should be detectable at high latitudes, $\sim100\MHz$ frequencies and sub-degree scales. The signal may have already been detected by CMB anisotropy studies, at $\sim 10\GHz$ frequencies and $\theta <1\de$ scales \cite{WaxmanLoeb2000}, and in quiet regions of $\sim \GHz$ high resolution surveys at $1\marcmin-10\marcmin$ scales \cite{Keshet04a}. It will probably be easiest to identify the signal at arcminute scales, by modelling point sources and cross-correlating the maps with known tracers of large-scale structure.

Future detection of radio emission from intergalactic shocks will have important implications on our understanding of cosmology and astrophysics. Detection of the signal will provide the first identification of intergalactic shocks, revealing the underlying cosmological flows and providing a test for structure formation models. The signal, in particular when combined with \gama-ray detection, will provide a measure of the unknown magnetic fields in the intergalactic medium. 
Although non-trivial for interpretation, such a measure of the magnetic field may provide insight into the unknown processes leading to IGM magnetization. The signal may confirm the existence of the undetected warm-hot intergalactic medium, and provide a tracer for its distribution. 

Synchrotron emission from intergalactic shocks is correlated with the large-scale structure of the low-redshift ($z<1$) universe, tracing young galaxy clusters and filaments. The signal could thus account for some observed features of the radio signature of galaxy clusters, namely radio halos and radio relics. Extended accretion shocks could contribute to radio halos (e.g. to the large radio halo of the Coma super-cluster, see \cite[and the references therein]{Thierbach03}), whereas localized shocks (e.g. where galaxy filaments channel large amounts of gas into the cluster regions) may be responsible for some radio relics observed at the outskirts of clusters (such as the prototype relic found in the Coma super-cluster, 1253+275 \cite{Ensslin98}), and perhaps also for some of the anomalous features observed in radio halos (e.g. in the unrelaxed clusters described in \cite{Govoni04}). 

Radio emission from intergalactic shocks is an important source of
contamination for other radio signals, such as the low frequency CMB,
Galactic synchrotron fluctuations on sub-degree scales, and competing
extragalactic radio signals such as bremsstrahlung from \Lya clouds, and 21
cm tomography. 
Emission from intergalactic shocks should be taken into account when evaluating the propagation of ultra-high energy photons with energies above
$10^{19}\eV$, where the effect of the radio background on the transparency of the universe is stronger than the effect of the CMB.

\end{document}